\documentclass[final]{svjour3}
\usepackage{siunitx}
\usepackage{graphicx}
\usepackage{rotating}
\usepackage{amssymb}
\usepackage{mathptmx}
\usepackage[numbers]{natbib}
\makeatletter
\journalname{Journal of Low Temperature Physics}

\bibpunct{[}{]}{,}{n}{}{,}

\begin{document}

\newcommand{\hdblarrow}{H\makebox[0.9ex][l]{$\downdownarrows$}-}
\title{Design of The Kinetic Inductance Detector Based Focal Plane Assembly for The Terahertz Intensity Mapper}

\author{L.-J.~Liu\textsuperscript{1} \and R.M.J.~Janssen\textsuperscript{1,2} \and C.M.~Bradford\textsuperscript{1,2} \and S.~Hailey-Dunsheath\textsuperscript{1,2} \and J.~Fu\textsuperscript{3} \and J.P.~Filippini\textsuperscript{3} \and J.E.~Aguirre\textsuperscript{4} \and J.S.~Bracks\textsuperscript{4} \and A.J.~Corso\textsuperscript{4} \and C.~Groppi\textsuperscript{5} \and J.~Hoh\textsuperscript{5} \and R.P.~Keenan\textsuperscript{6} \and I.N.~Lowe\textsuperscript{6} \and D.P.~Marrone\textsuperscript{6} \and P.~Mauskopf\textsuperscript{5} \and R.~Nie\textsuperscript{3} \and J.~Redford\textsuperscript{1} \and I.~Trumper\textsuperscript{7} \and J.D.~Vieira\textsuperscript{3}}

\institute{\textsuperscript{1}California Institute of Technology, Pasadena, CA 91125, USA;
\email{lliu@caltech.edu}\\
\textsuperscript{2}Jet Propulsion Laboratory, California Istitute of Technology, Pasadena, CA 91109, USA\\
\textsuperscript{3}University of Illinois at Urbana-Champaign, Urbana IL 61801, USA\\
\textsuperscript{4}University of Pennsylvania, Philadelphia, PA 19104 USA\\
\textsuperscript{5}Arizona State University, Tempe, AZ 85281, USA\\
\textsuperscript{6}University of Arizona, Tucson, AZ 85721, USA\\
\textsuperscript{7}Intuitive Optical Design Lab LLC, Tucson, AZ 85701, USA}

\maketitle

\begin{abstract}

We report on the kinetic inductance detector (KID) array focal plane assembly design for the Terahertz Intensity Mapper (TIM). Each of the 2 arrays consists of 4 wafer-sized dies (quadrants), and the overall assembly must satisfy thermal and mechanical requirements, while maintaining high optical efficiency and a suitable electromagnetic environment for the KIDs. In particular, our design manages to strictly maintain a 50 $\mathrm{\mu m}$ air gap between the array and the horn block. We have prototyped and are now testing a sub-scale assembly which houses a single quadrant for characterization before integration into the full array. The initial test result shows a $>$95\% yield, indicating a good performance of our TIM detector packaging design.


\end{abstract}

\section{Introduction}

More than half of the star formation history of the Universe has taken place in dust-obscured regions~\cite{MH2001,GL2005}. In these dusty galaxies, radiation of stars is absorbed by dust and re-radiated in infrared (IR) wavelengths. Therefore, studying the IR universe is crucial to discover the abundance of information dust obscured galaxies can provide on the nature of star formation and galaxy evolution.

The Terahertz Intensity Mapper (TIM) is a balloon-borne line intensity mapping experiment designed to unravel the formation and evolution of galaxies around the peak of cosmic star formation (0.5$<$z$<$1.7)~\cite{JV2019}. TIM will perform spectroscopic measurements of redshifted far-IR spectral lines, most importantly the redshifted 158 $\mathrm{\mu m}$ line of singly ionized carbon ([CII]), to provide an unbiased dust-immune census of star formation history. TIM will apply line intensity mapping to measure the total line intensity of all galaxies in the volume set by its spectral and spatial resolution~\cite{BU2014}. This novel methodology will enable TIM to determine the three-dimensional structure of our star forming universe.

\section{TIM Instrument}

The TIM payload will be mounted on a balloon, whose flight is planned to be in Antarctica at an altitude of 37 km. The TIM telescope design is lightweight and compact, with low overall emissivity~\cite{JV2019}. A 2-meter mirror is fabricated as the primary of the entire TIM optics. TIM employs two R $\sim 250$ long-slit grating spectrometers covering $240-317$ $\mathrm{\mu m}$ (short wavelength band; SW) and $317-420$ $\mathrm{\mu m}$ (long wavelength band; LW). Each band is equipped with a focal plane unit containing 4 wafer-sized subarrays (termed quadrants) of aluminum kinetic inductance detectors (KIDs). The entire array focal plane packages are cooled to 250 mK using a multi-stage liquid $^3$He sorption fridge (so called He-10 fridge).

\section{Design of Individual Detectors}

KIDs are thin-film superconducting resonators which absorb incident radiation and respond by a change in their resonance frequency and quality factor~\cite{JZ2012,PD2003}. TIM implements a lumped-element KID design made from 30 nm aluminum, where each micro-resonator consists of an inductive absorber and an interdigitated capacitor. The capacitor's geometry can be tuned to give each detector on the chip a unique resonance frequency, and so it naturally provides the advantage of frequency-division multiplexing. Every KID is capacitively coupled to the microstrip feed line. Such a lumped-element KID design enables a large array of detectors to be read out on a single RF/microwave circuit~\cite{JV2019}. These detectors are designed to be photon-noise-limited under an optical loading of 100 fW at an operational temperature of 250 mK. A detailed description of the individual KID design as well as a measurement of KIDs' optical sensitivity can be found in Janssen et al.~\cite{RJ2021}.

\begin{figure}[htbp]
\begin{center}
\includegraphics[width=0.75\linewidth, keepaspectratio]{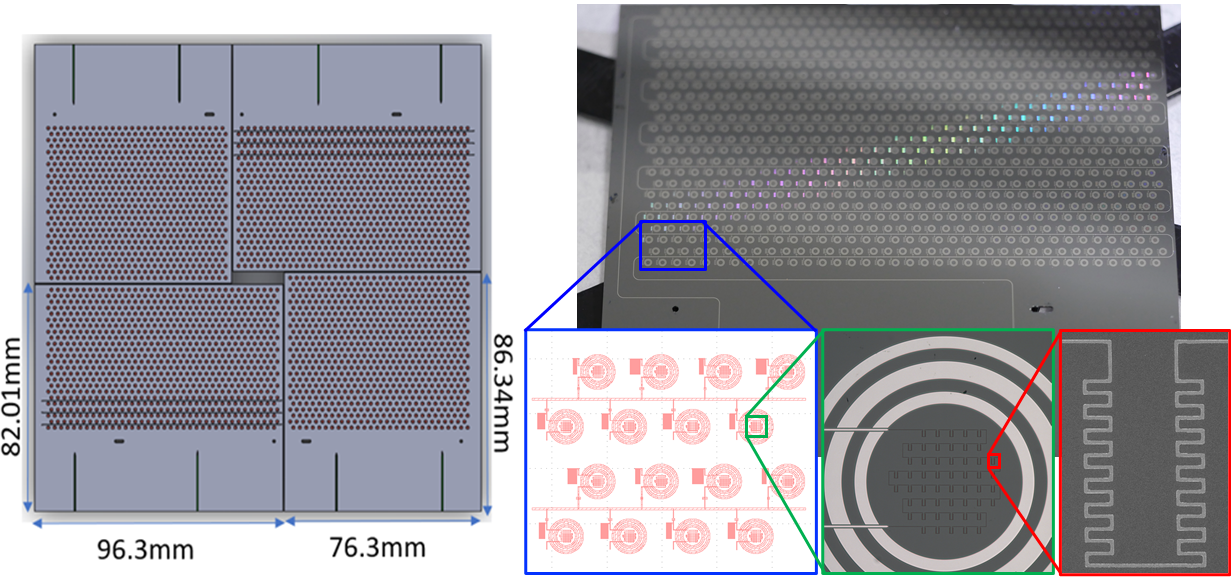}
\caption{Left: Layout of the LW TIM detector wafers. The 2 unique quadrant designs are named as main array ($96.3 \times 82.01$ mm) and secondary array ($76.3 \times 86.34$ mm), each of which is present twice in the full array. Right: The picture of one fabricated LW main array. From bottom left to right, the zooms show a unit cell of 16 pixels, the full inductive absorber, and the ``chain-link'' absorber unit cell design from which the inductor is built up.}\label{fig1}
\end{center}
\end{figure}

In every wafer-sized quadrant subarray of the full array, the KIDs are hex-packed with 2.3 mm spacing and optically coupled via a matching arrays of direct-machined feedhorns in an aluminum horn block. The geometry of each KID's interdigitated capacitor is tuned to assign a unique resonance frequency, and together all KIDs are able to fill the entire 0.5 GHz readout bandwidth. This enables the quadrant to be read out using a ROACH2-based readout system. Fig.~\ref{fig1} (right) demonstrates this fabricated LW quadrant wafer with zooms on the details of detector design.

High absorption efficiency in the far-IR requires (1) a proper design of the absorber geometry and (2) a small and steady gap between the array chip and the feedhorn horn block to minimize optical crosstalk. For the first requirement, we implement a novel ``chain-link'' (CL) absorber design that offers excellent performance. The bottom right zoom of Fig.~\ref{fig1} demonstrates this novel CL design. We conduct a simulation project using ANSYS-HFSS\footnote{https://www.ansys.com/products/electronics/ansys-hfss}, and the optimization result indicates that this CL design reaches an in-band absorption efficiency over $\sim90\%$ in both linear polarization modes~\cite{RN2021}. In order to embrace the second aspect, we first quantify the optical crosstalk in various air gaps. The HFSS simulation on the TIM's first-generation absorber design demonstrates the increase of the air gap from 40 $\mathrm{\mu m}$ to 80 $\mathrm{\mu m}$ causes the radiation leakage to increase by $\sim10$ times and significantly increases the optical crosstalk with neighbor pixels~\cite{RN2020}. A detailed description of these simulations and optimizations are presented in Nie et al.~\cite{RN2021,RN2020}. Our design approach therefore requires a well-controlled 50 $\mathrm{\mu m}$ spacing between the array chip and the horn block in a package that must survive thermal cycling. As a result, this architecture utilizes the horn block surface facing the silicon as the ground of the microstrip readout line.

\section{Full Array Detector Packaging Design}

\begin{figure}[htbp]
\begin{center}
\includegraphics[width=0.85\linewidth, keepaspectratio]{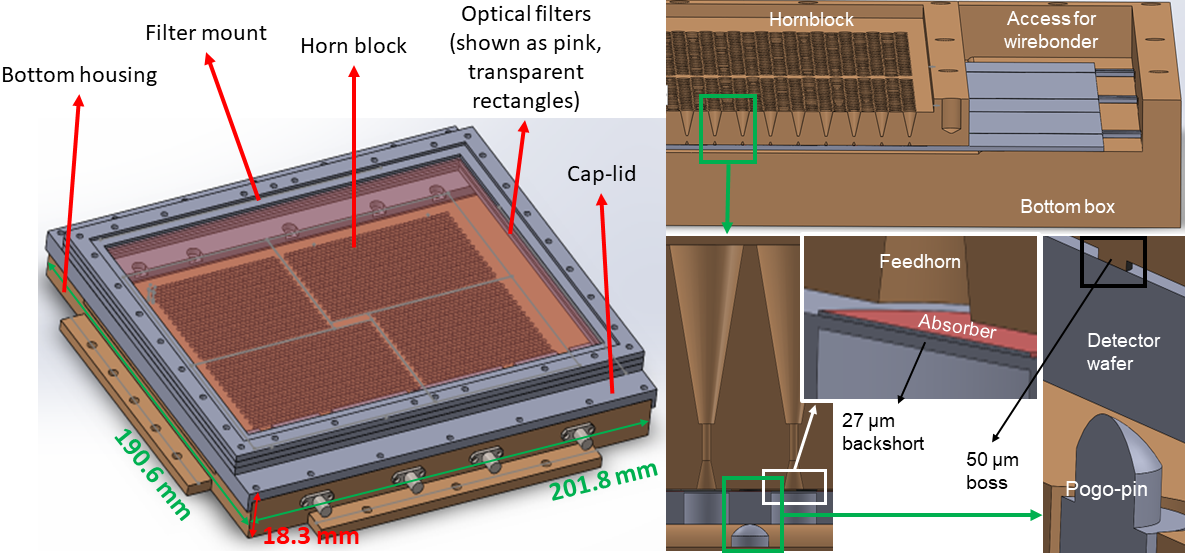}
\caption{Left: The solid model of the LW full array detector packaging. Note that the optical filters are mounted above the horn block and shown as pink, transparent parallelograms. Right: A detailed description of the detector wafer mount inside the package. Bottom right: A 50 $\mathrm{\mu m}$ boss is located on the detector wafer (highlighted in the black rectangle) and a matching spring-loaded pin pushes against the wafer.}\label{fig2}
\end{center}
\end{figure}

We present the LW focal plane unit design as representative, and note that the SW design will be nearly identical. Fig.~\ref{fig2} (left) shows the solid model of the LW full array packaging. From top to bottom, the components are (1) a filter assembly which allows stacking 2 optical filters (shown as pink, transparent parallelograms), (2) a cap-lid which presses down the wire bond section of wafer using spring-loaded pins, (3) a horn block with feedhorns for all 4 quadrant subarrays, and (4) a bottom housing which mounts the same type of spring-loaded pins as well as 2 sets of 4 SMA connectors on 2 sides. Each TIM focal plane unit contains 4 quadrants using 2 unique designs that together cover the 64 spectral by 51 spatial pixels of the full array. As shown in Fig.~\ref{fig1} (left), the quadrants are positioned so that the necessary gaps between quadrants are staggered and no spatial position or spectral channel is lost for the full array. Through-holes etched in the wafer provide in-plane alignment via matching guide pins in the horn block.

We enable a steady 50 $\mathrm{\mu m}$ air gap with an array of spring-loaded pins pushing the chip against a matching array of precision-machined bosses on the bottom of the horn block. Details of this scheme are demonstrated in Fig.~\ref{fig2} (right), where one example is shown in the bottom right picture. This approach offers compliance to thermal deformations on cooldown, while avoiding any front-side circuitry. Specifically, the positions and patterning of these bosses are arranged so that none of them will damage the detectors and microstrip line due to the position shift caused by the differential thermal contraction between the aluminum and the silicon during cooldown. Also, this ``clamping'' scheme spans the full array to provide even forces to all 4 quadrants.

\begin{figure}[htbp]
\begin{center}
\includegraphics[width=0.7\linewidth, keepaspectratio]{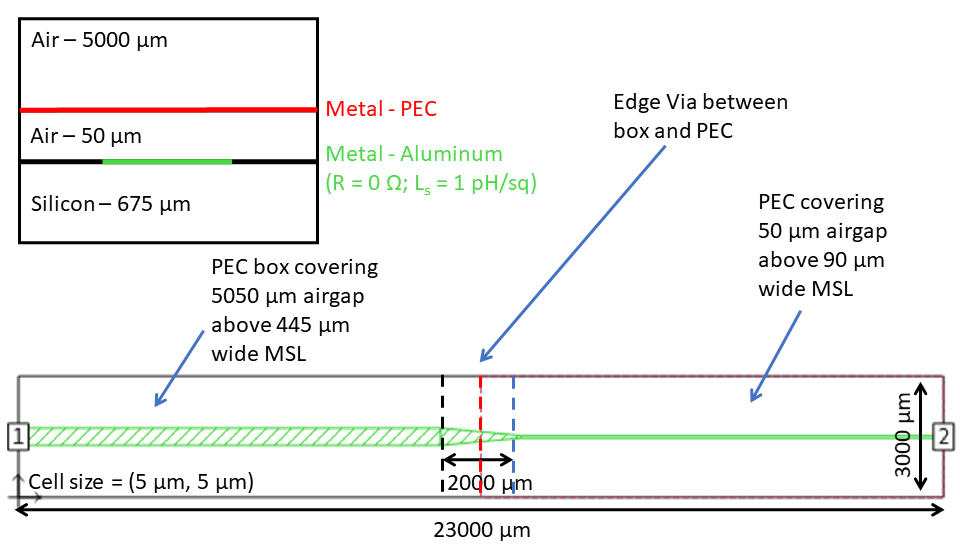}
\caption{The setup of an electromagnetic simulation project using commercial software SONNET to determine the feasibility of the microstrip line transitioning out from under the horn block. The top left part shows the medium around microstrip line, and the bottom part shows the configuration of the line transition.}\label{fig3}
\end{center}
\end{figure}

\begin{figure}[htbp]
\begin{center}
\includegraphics[width=0.95\linewidth, keepaspectratio]{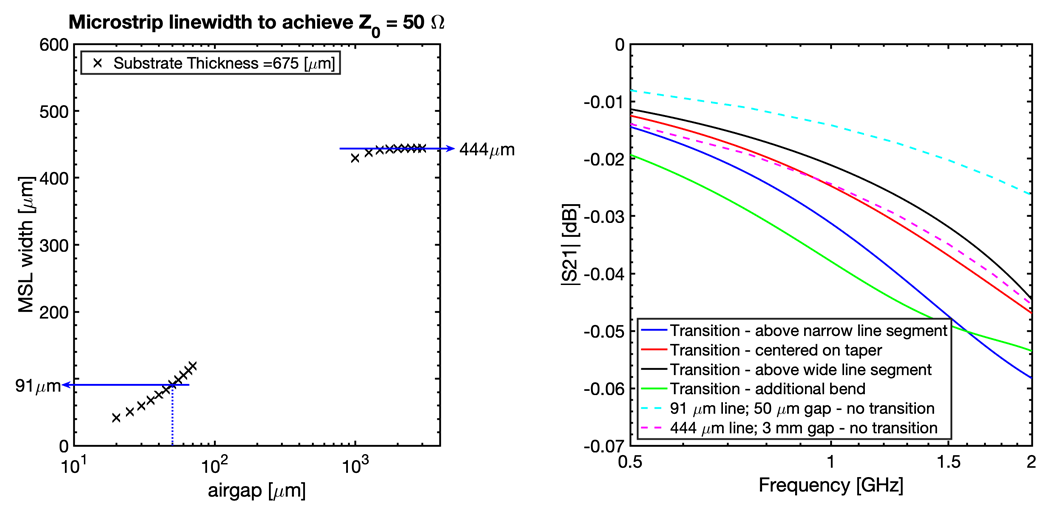}
\caption{Left: The change of readout line width with respect to the air gap for matching a \SI{50}{\ohm} impedance. Below the horn block is 50 $\mathrm{\mu m}$ gap, and so our SONNET simulation indicates a 91 $\mathrm{\mu m}$ line width. The value of microstrip line width saturates at 444 $\mathrm{\mu m}$, indicating that this particular line width should match \SI{50}{\ohm} in the bonding region. Right: Study of the signal loss in the transition structure using the SONNET setup shown in Fig.~\ref{fig3}. The exact location of the transition is varied across the taper as indicated by the blue, red, and black lines in both figures. The simulation result shows that in any case the loss in this transition is $<0.1$ dB.}\label{fig4}
\end{center}
\end{figure}

Due to the proximity of the horn block it will be the dominant ground plane within its covered region. Therefore, we should determine the geometry of the microstrip line for impedance matching. We have performed SONNET simulations\footnote{https://www.sonnetsoftware.com/} to determine the required line widths to achieve a $Z_0 =$ \SI{50}{\ohm} feedline. The simulation consists of a 2 by 2 mm box with walls, ceiling and bottom made from a perfect electric conductor. From bottom to top the simulated layers consist of 675 $\mathrm{\mu m}$ of intrinsic silicon (set by the wafer thickness), superconducting aluminum $(L_s = 1$ pH/sq; $R_{dc}=R_{rf}=$ \SI{0}{\ohm}) and a vacuum layer of which the height is varied between 20 and 2500 $\mathrm{\mu m}$. The width of the line in the aluminum is varied in steps of 5 $\mathrm{\mu m}$. Fig.~\ref{fig4} (left) shows the line width for which a \SI{50}{\ohm} impedance is achieved as a function of the air gap height. For the 50 $\mathrm{\mu m}$ air gap the optimal width is 91 $\mathrm{\mu m}$. It is worth noting the impedance of the line is robust against variations in both the air gap height and line width; a deviation of $\sim 10$\% in either results in $\sim 5$\% change in $Z_0$. At large air gaps, representative of the bonding section, the ground of the microstrip is the bottom of the wafer (where the aluminum of the backshort is present) and the optimum line width saturates at 444 $\mathrm{\mu m}$.

As the readout line microstrip emerges from the optically-active region, the stripline widens to transition smoothly from being grounded by the horn block to being grounded through the silicon to the bottom of the chip. A key feature of the packaging design, this transition not only involves a sharp change in the air gap, but also a change in the location of the dominant ground. Fig.~\ref{fig3} demonstrates the setup in SONNET used to simulate this transition. The feedline is tapered over a length of 2 mm, narrowing from 444 to 90 $\mathrm{\mu m}$. Fig.~\ref{fig4} (right) shows the transmission across the transition, for 3 locations of the air gap step (black, red, blue -- colors matched to the locations indicated in Fig.~\ref{fig3}). In addition, one simulation has been done in which a 90 degree bend with an inner radius of 1 mm is placed 10 mm after the transition, shown in Fig.~\ref{fig4} (right, green line), as is common in the full array design. For reference, Fig.~\ref{fig4} (right) also shows the loss of a single 444 (dashed magenta line) or 90 $\mathrm{\mu m}$ (dashed cyan line) wide line with appropriate air gap over the 23 mm distance used in the transition simulation. Even though the transition increases the transmission loss somewhat, at an absolute value of $<0.1$ dB should still be fully negligible and the transition is not expected to be an issue during operation.

\section{Assembly and First Test of Quadrant Array Packaging}

\begin{figure}[htbp]
\begin{center}
\includegraphics[width=0.85\linewidth, keepaspectratio]{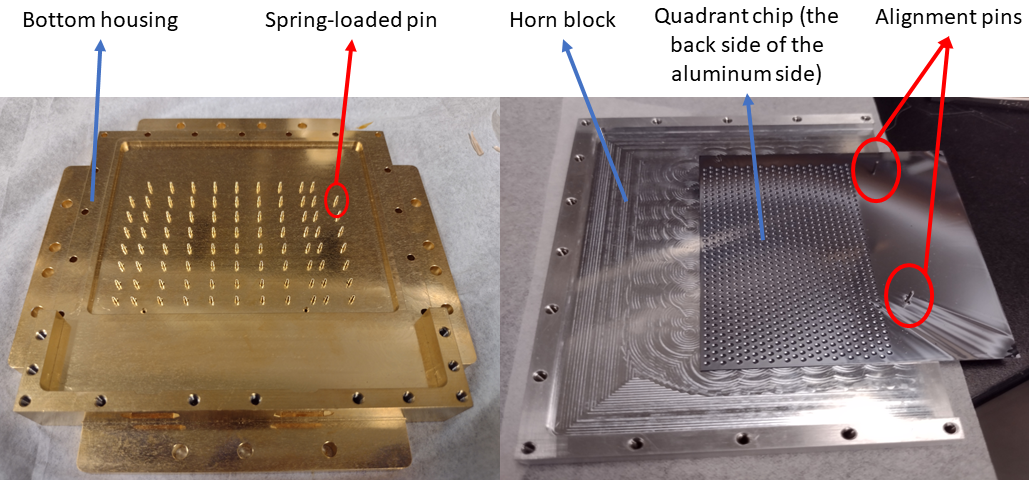}
\caption{The full array packaging design is scaled down as a quadrant package for testing individual quadrant array chip. Left: an array of spring-loaded pins are glued on the bottom housing using silver epoxy H20E. Right: The first step of assembling the focal plane unit. Two alignment pins are epoxied at the back side of a dummy horn block, as highlighted in red circles. We face the aluminum side of wafer to 50 $\mathrm{\mu m}$ bosses (hidden here), and then we gently slide the chip through pins and rest it on the bosses.}\label{fig5}
\end{center}
\end{figure}

We will initially test and characterize the quadrant chips individually before integration into the full array. Fig.~\ref{fig5} shows 2 parts in the quadrant assembly we have built, using the same microstrip architecture as will be used in the full array. The left panel shows an array of spring loaded pins bonded on the bottom housing using silver epoxy H20E, and the right part shows a quadrant array chip resting on the back of horn block with its location constrained by 2 alignment pins (highlighted in red circles) interfacing with a pin and slot in the silicon. Initial cryogenic tests and measurements with this system are underway now. This test uses a pulse-tube pre-cooled triple-stage He sorption cooler with a base temperature of 213 mK.

\begin{figure}[htbp]
\begin{center}
\includegraphics[width=0.85\linewidth, keepaspectratio]{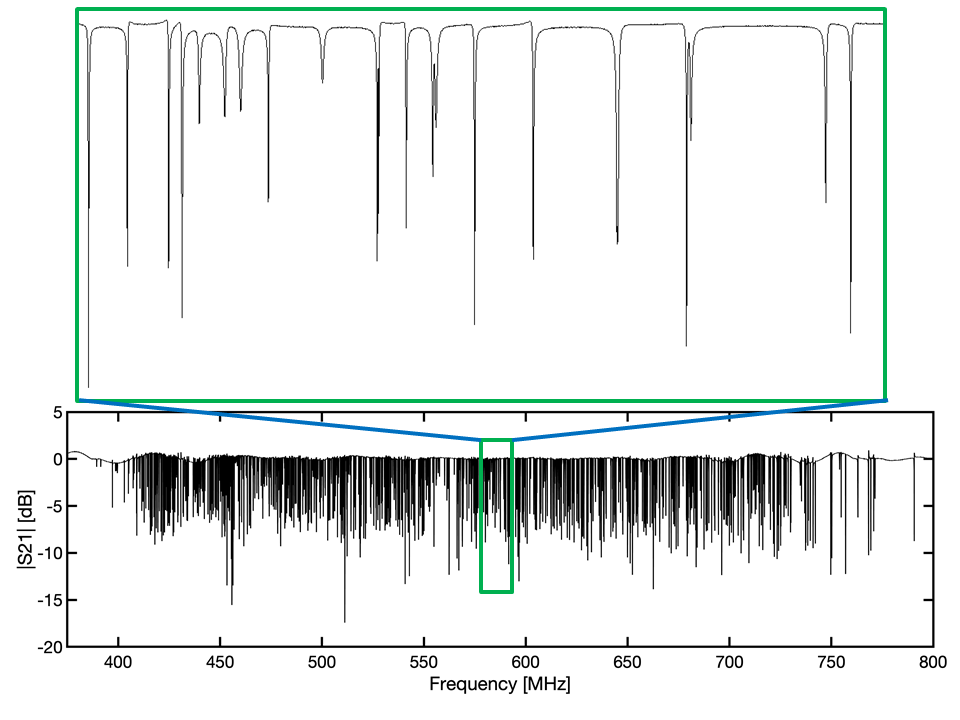}
\caption{The VNA sweep of the 864 pixel LW quadrant array from 360 MHz to 800 MHz and a zoom picture of a segment of resonator responses at $\sim 580$ MHz. We discover 823 resonators across the readout frequency band, and so over 95\% of the total detectors are present.}\label{fig6}
\end{center}
\end{figure}

We are conducting the dark measurement for the initial test, and so the horn block is replaced with a solid lid without any feedhorns, but stress-relieved and with the carefully-machined 50 $\mathrm{\mu m}$ bosses. Fig.~\ref{fig6} shows a preliminary vector network analyzer (VNA) scan of this system. We identify 823 resonators out of 864 designed, indicating a high yield (95.25\%). The resonator frequencies are within range of their design values, and the quality factors are set by the coupling as designed ($Q_c \sim 10^{5}$ ), indicating that our electromagnetic design of the system is realistic, and that the mechanical architecture is robust.

\section{Summary and Future Prospects}

We have designed the full KID array package for TIM, which utilizes a novel approach to (1) maintain the 50 $\mathrm{\mu m}$ air gap accurately and reliably, (2) provide compliance to thermal deformations on cooldown, as well as (3) smoothly transit the ground along the microstrip readout line. The quadrant-level package which demonstrates this novel approach is built and under test now. A preliminary VNA test has demonstrated a $>95\%$ yield of the 864 pixel quadrant chip, which validates the viability of the packaging design. This yield will provide excellent performance when deployed in TIM, which is a spatial-spectral multiplexing instrument for which the science degrades gracefully as pixels are removed. We are conducting further measurements to unambiguously identify each resonance with a cryogenic LED array, allowing us to eliminate collisions via capacitor trimming~\cite{XL2017}. Subsequently, we will confirm that the responsivity and sensitivity in this 864-pixel quadrant array match that found in the small subarrays (Janssen et al.~\cite{RJ2021}).

\begin{acknowledgements}
TIM is supported by NASA under grant 80NSSC19K1242, issued through the Science Mission Directorate. R.M.J. Janssen is supported by an appointment to the NASA Postdoctoral Program at the NASA Jet Propulsion Laboratory, administered by Oak Ridge Associated Universities under contract with NASA. Part of this research was carried out at the Jet Propulsion Laboratory, California Institute of Technology, under a contract with the National Aeronautics and Space Administration (80NM0018D0004).
\end{acknowledgements}

\end{document}